\documentclass[11pt,a4paper]{article}
\usepackage[latin9]{inputenc}
\usepackage{graphicx}

\makeatletter

\pdfpageheight\paperheight
\pdfpagewidth\paperwidth

\providecommand{\tabularnewline}{\\}

\pdfoutput=1 

\usepackage{jinstpub}

\title{Data Acquisition System for the CALICE AHCAL Calorimeter}

\author[a,b]{J. Kvasnicka}

\affiliation[a]{Institute of Physics of the Czech Academy of Sciences,\\Na Slovance 2, 18221 Prague 8, Czech Republic}
\affiliation[b]{DESY Deutsches Elektronen-Synchrotron,\\Notkestrasse 85, 22607 Hamburg, Germany}
\emailAdd{kvas@fzu.cz}

\abstract{The data acquisition system 
(DAQ)
for a highly granular analogue hadron 
calorimeter 
(AHCAL) 
for the future International Linear Collider 
is presented. 
The developed DAQ chain has several stages of aggregation and scales up to  
8 million channels foreseen for the AHCAL detector design. 
The largest aggregation device, 
Link Data Aggregator, has 
96 HDMI connectors, four Kintex7 FPGAs and a central Zynq 
System-On-Chip. Architecture and performance results 
are shown in detail. Experience from DESY testbeams with 
a small detector prototype consisting of 15 detector layers are shown.
}

\keywords{Calorimeters, 
Data acquisition concepts, 
Data acquisition circuits,
Front-end electronics for detector readout}

\collaboration[c]{on behalf of the CALICE collaboration}

\proceeding{TWEPP 2016 - Topical Workshop on Electronics for Particle Physics\\
26 September 2016 - 30 September 2016\\
Karlsruhe Institute of Technology (KIT), Germany}

\usepackage{lineno}
\usepackage[binary-units]{siunitx}

\makeatother

\begin{document}

\maketitle \flushbottom

\section{Introduction }

The international CALICE collaboration \cite{calice} is developing
calorimeters for detectors at future linear electron-positron colliders
with a very high granularity, which is essential for the particle
flow reconstruction algorithm \cite{pfa} to achieve the best jet
energy resolution. The AHCAL (Analog Hadron CALorimeter) group is
developing a steel sandwich hadron calorimeter option, that uses $\SI[product-units = power]{3 x 3 x 0.3}{cm}$
scintillating tiles for light conversion and SiPM (Silicon Photo-Multipliers)
detectors for reading out each tile individually. This granularity
corresponds to in total 8 million channels in the  AHCAL barrel and
endcaps.

The data acquisition (DAQ) concept needs to cope with such granularity
by defining aggregation stages. The DAQ concept \cite{daq_concept,DAQ_Goodrick}
that was proposed for calorimeters for ILD (International Large Detector,
\cite{tdr_ild}) is scalable to the final detector system. The first
generation CALICE prototypes used the same DAQ system in beam tests.
Since then, the DAQ systems have been individually adapted for the
technological prototypes. The current AHCAL DAQ, which utilizes data
concentration electronics, is also used by the scintillator-based
ECAL (Electromagnetic Calorimeter) prototype.

The AHCAL DAQ serves two distinct purposes: it provides a) a stable
DAQ usable for tests of the detector prototypes in the lab and in
beam tests; b) a technological demonstrator of a DAQ, that can be
scaled to the final detector system and ILC (International Linear
Collider) timing.

\section{Architecture }

The architecture reflects the timing of the ILC  accelerator and
power constraints given by the ILD. The timing requirements are specific
to the ILC, where the collisions take place in a \SI{\sim1}{ms} train
of bunches spaced \SI{\sim300}{ns} apart, followed by \SI{199}{ms}
idle time. The power constraints reflect the lack of active cooling
within detector volume in the AHCAL design \cite{tdr_ild}, where
the temperature difference between the cooled sides and the center
limits the average power dissipation of a single channel to \SI{25}{\micro\watt}
per channel, when a $\SI[product-units = power]{3 x 3 x 0.3}{cm}$
geometry is considered.

\begin{figure}
\centering{}\includegraphics[width=10.5cm]{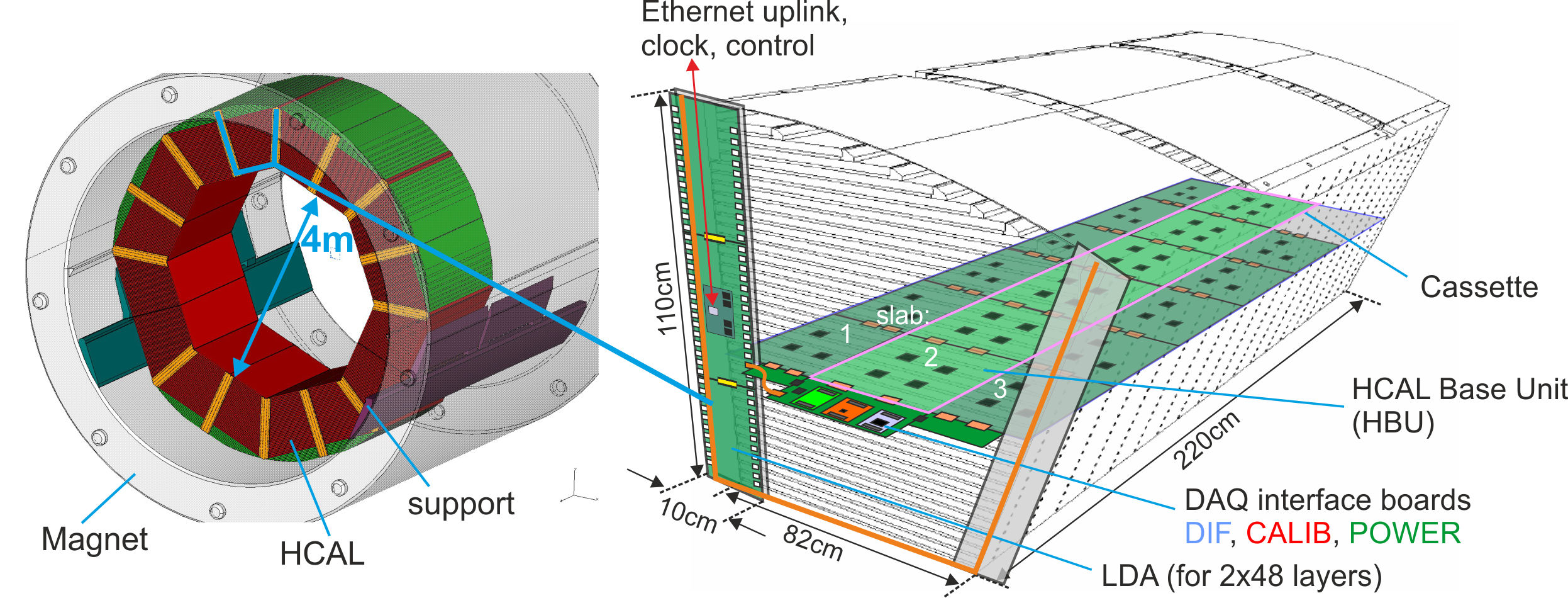}\hfill{}\includegraphics[width=4.5cm]{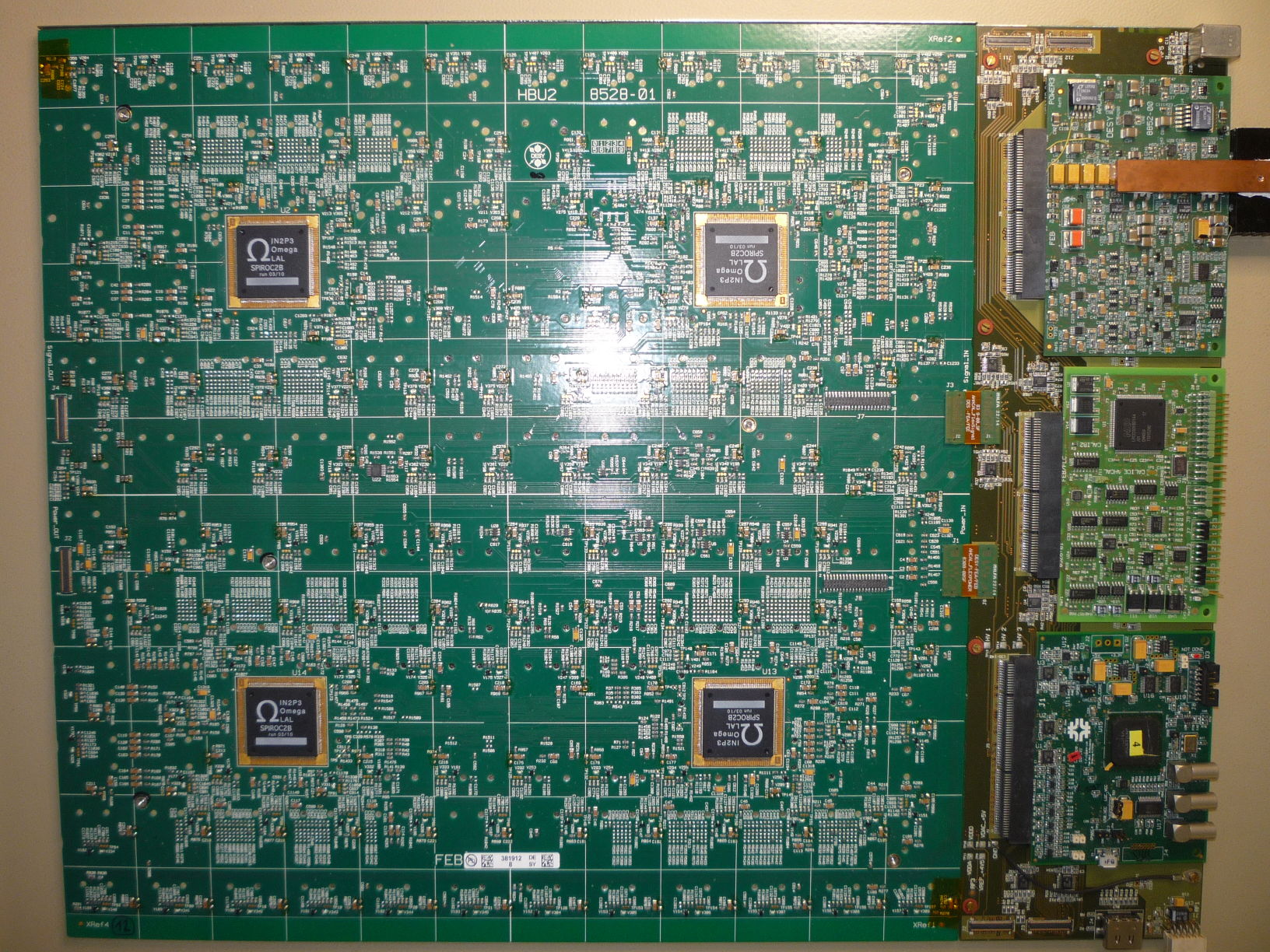}\caption{\label{fig:The-AHCAL}The AHCAL barrel geometry with a single wedge
detail showing 1/16 of the barrel (left), and  one AHCAL Base Unit
(HBU) connected to the DAQ interface electronics (right).}
\end{figure}

The electronics geometry is limited by a \SI{5.4}{mm} gap between
absorber material, which has to contain a \SI{3}{mm} thick plastic
scintillating tile with SiPM detector, a \SI{0.7}{mm} thinned PCB
and \SI{1.5}{mm} for SMD components and connectors. The active electronics
and detection layer spans \SI{2.2}{m} along the barrel (figure \ref{fig:The-AHCAL}
(left)), and is segmented into HBUs (AHCAL base units, figure \ref{fig:The-AHCAL}
(right)) with \SI[product-units=power]{36x36}{cm} size.

\begin{figure}
\begin{centering}
\includegraphics[width=8cm]{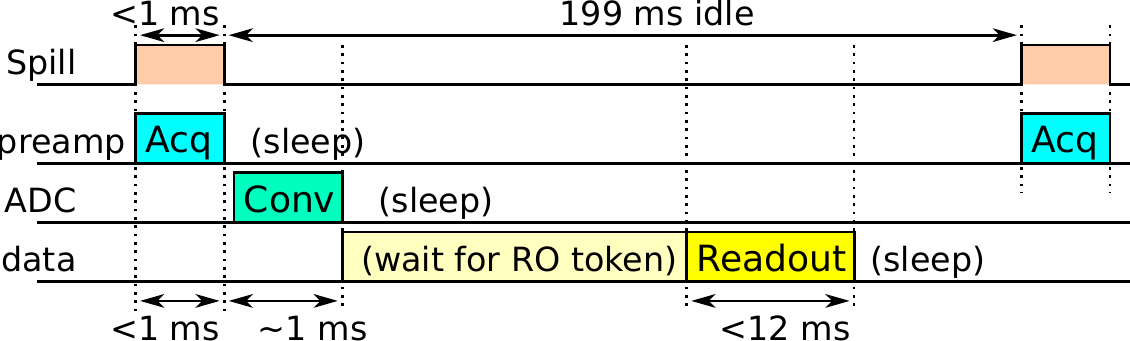}
\par\end{centering}
\caption{\label{fig:Acquisition-phases}Acquisition phases and power sequence
in ILC timing scenario for a single ASIC.}
\end{figure}

To achieve low power dissipation, the acquisition cycle is split in
3 phases, shown in figure \ref{fig:Acquisition-phases}: 
\begin{enumerate}
\item The \emph{acquisition phase}, where the electrical signal is saved
in the form of charge in analog memory cells in the very front-end
ASICs. A zero suppression is applied by a self-triggered operation.
\item The \emph{conversion phase}, where all stored analogue signals are
converted to digital signals by a single ADC integrated into the very
front-end ASIC.
\item The\emph{ readout phase}, where the data is shifted out from the ASICs
further in the DAQ chain. The data bus is shared among 12 ASICs in
the readout chain.
\end{enumerate}
Each phase has independent power control, where the power is enabled
only for the preamplifiers in the acquisition phase, the ADC is enabled
only in the conversion phase and the digital part is enabled only
in the readout phase.

\subsection{Hierarchy}

\begin{figure}
\begin{centering}
\includegraphics[width=10cm]{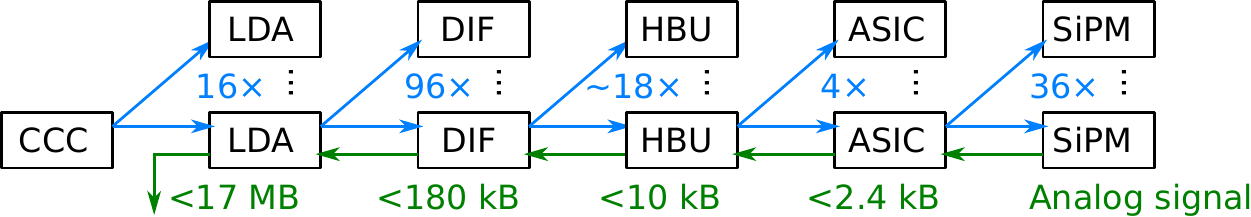}
\par\end{centering}
\caption{\label{fig:Maximum-data-volume}Aggregation factors and maximum data
volume in a single acquisition cycle.}
\end{figure}

The AHCAL DAQ architecture adopted the aggregation hierarchy from
the DAQ concept \cite{daq_concept}, but the hardware was developed
from scratch. The aggregation factors are shown in figure \ref{fig:Maximum-data-volume}.
The following list describes the individual components in the DAQ
chain:
\begin{description}
\item [{SPIROC}] is an ASIC developed by the Omega group \cite{spiroc},
which has 36 input channels, an 0SCA (Switched Capacitor Array) for
storing up to 16 events in analog form and an integrated 12-bit ADC.
Charge (ADC) and Time (TDC) information is stored for each channel
independently. Only signals passing a programmable threshold are stored.
\item [{HBU~(AHCAL~Base~Unit)}] is a $\SI[product-units = power]{36 x 36}{cm}$
large board developed at DESY \cite{mark_engineering_prototype},
which has 4 SPIROCs organized into 2 readout chains. HBUs can be connected
to each other via flat connectors, forming a slab of 6 HBUs up to
\SI{2.2}{m} long as shown in figure \ref{fig:The-AHCAL} (left).
\item [{DIF~(Detector~InterFace)}] is the first aggregation component,
which controls and reads out up to 3 slabs, which corresponds to 18
HBUs or 72 ASICs. The data is sent out via HDMI connectors with a
custom serial protocol. The latest version contains a Zynq-7020 SoC
(System-on-Chip), an FPGA with an embedded dual-core ARM processor
\cite{xilinx_zynq}.
\item [{LDA~(Link~Data~Aggregator)}] reads out all DIFs from 2 wedges
of the AHCAL barrel (figure \ref{fig:The-AHCAL} (left)) from 96 HDMI
inputs and sends the data directly to a computer via GbE (Gigabit
Ethernet). Details are shown in section \ref{subsec:Link-Data-Aggregator}.
\item [{CCC~(Clock\_and\_Control\_Card)}] delivers a beam clock to the
LDA and starts the acquisition cycle with the presence of a spill.
It synchronizes the clock in common beam tests of different detectors.
During beam tests it also delivers a delayed trigger validation signal.
\end{description}

\subsection{\label{subsec:Link-Data-Aggregator}Link Data Aggregator (LDA)}

\begin{figure}
\begin{centering}
\includegraphics[width=4cm]{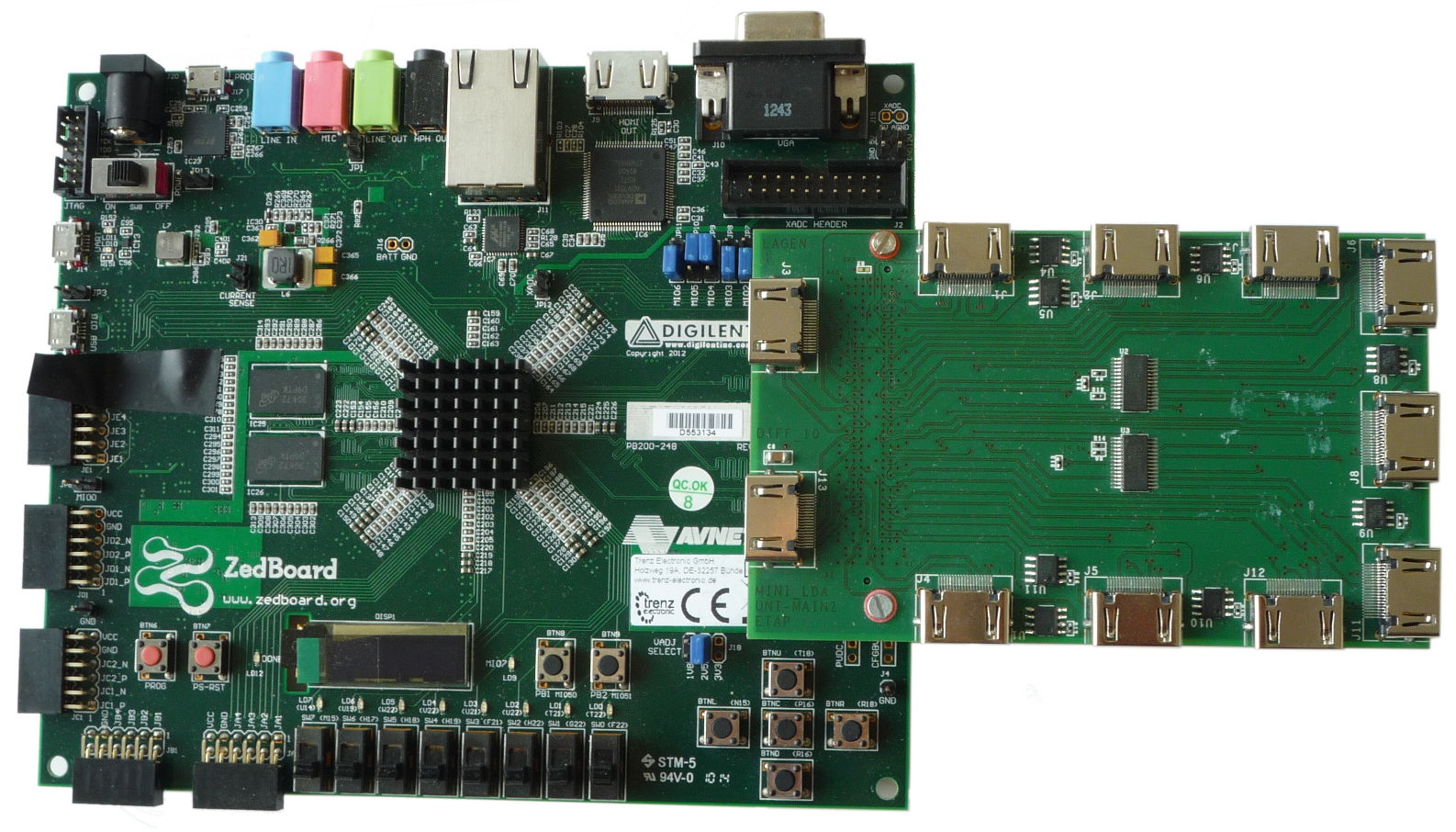}\hfill{}\includegraphics[width=10.5cm]{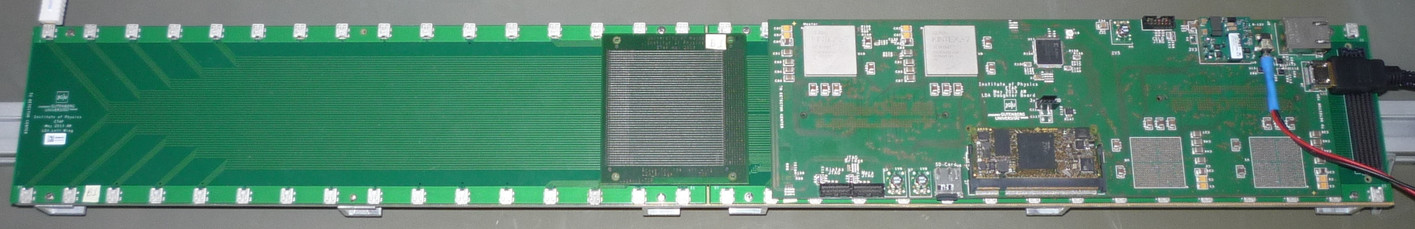}
\par\end{centering}
\caption{\label{fig:lda-form-factor}Two LDA form factors: a Mini-LDA, a 10-ports
mezzanine on Zedboard commercial FPGA board (left); a Wing-LDA with
2 FPGAs assembled and with one side wing, reaching only $2/3$ of
its design length (right).}
\end{figure}

The LDA hardware exists in two form factors, that have the same functionality
and are interchangeable. The \emph{Mini-LDA }(figure \ref{fig:lda-form-factor}
(left)) is a custom FMC-LPC (FPGA Mezzanine Card - Low Pin Count)
mezzanine with 10 HDMI connectors to connect to DIFs and 1 HDMI connector
to connect to CCC. The mezzanine is attached to a Zedboard, a commercial
development board \cite{zedboard} hosting a Zynq-7020 SoC \cite{xilinx_zynq}.
The \emph{Wing-LDA} (figure \ref{fig:The-AHCAL} (right)) has in total
96 micro-HDMI ports. The connector pitch matches the AHCAL layer spacing
in the barrel and the overall length of the board is \SI{110}{\centi\meter},
as shown in figure \ref{fig:The-AHCAL} (left). It consists of 3 passive
fan-out boards with 32 micro-HDMI connectors each, that connect to
an active daughter-board. The daughter-board hosts 4 Kintex-7 FPGAs
(xc7k160), each serving 24 ports from passive boards. Those FPGAs
are connected to a central commercial Mars ZX3 SoC module \cite{mars-zx3},
which hosts the same Zynq SoC as the Mini-LDA.

\begin{figure}
\begin{centering}
\includegraphics[width=1\textwidth]{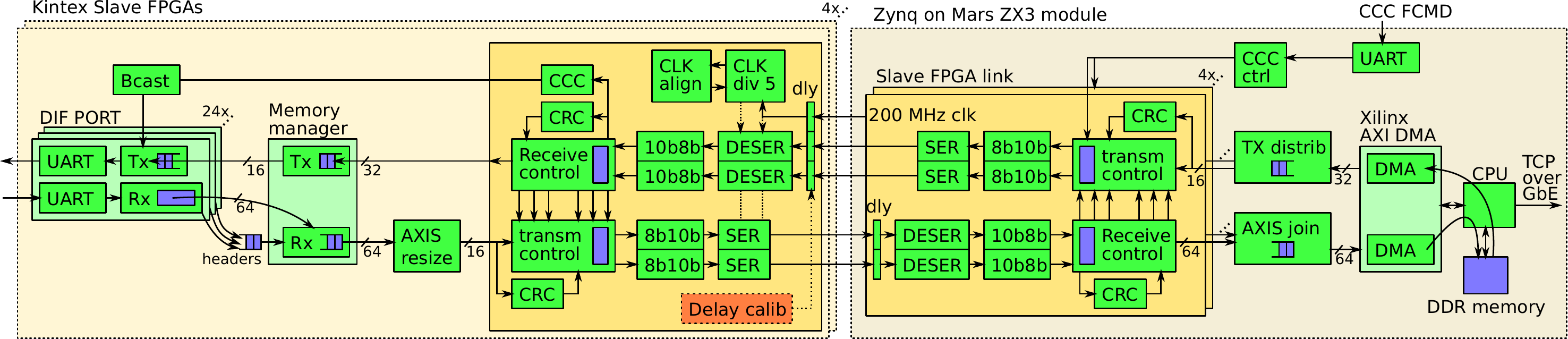}
\par\end{centering}
\caption{\label{fig:winglda-block-diagram}Wing-LDA data flow and block diagram.}
\end{figure}

The firmware of the both LDAs, Mini-LDA and Wing-LDA, share similar
source code, except for the Zynq $\leftrightarrow$ Kintex FPGA link,
which splits the design into 2 FPGA partitions in the Wing-LDA form
factor. The Wing-LDA FPGA firmware block diagram is shown in figure
\ref{fig:winglda-block-diagram}. The DIF sends 100-byte packet fragments
to the ``DIF PORT'' module, where the ASIC packets are built for
each SPIROC in the block memory in \SI{40}{\mega\hertz} clock domain.
Once a full ASIC packet is completed, the pointer is sent to the memory
manager, which moves the data from the memory of all ``DIF PORTS''
to a \SI{256}{\kibi\byte}  large buffer, from where it is transported
over the AXI-stream \cite{axi-stream} infrastructure. 

The ``FPGA link'' module has a circular buffer for up to 8 packets.
A CRC-16 check-sum is added for each packet. Two bytes are encoded
using the 8b10b code and serialized in \SI{200}{\mega\hertz} clock
domain in DDR mode, providing a dual \SI{400}{\mega\bit\per\second}
link in each direction. The check-sum of received packets is checked
and in case of error the packet is re-transmitted up to 3 times. Control
words (transmit succeed or fail, buffer full) are interleaved in the
stream via dedicated 8b10b control symbols. Fast commands from the
CCC (start and stop acquisition, synchronization) are also transmitted
as dedicated control symbols with highest priority, maintaining a
consistent command propagation delay.

Packets from the \SI{256}{\kibi\byte} AXI-stream FIFO are moved to
the DDR memory using an AXI DMA (Direct Memory Access) core in a simple
transaction mode (scatter-gather engine not used). Packet boundaries
are omitted in the DMA transfer in order to minimize the overhead
of initiating the transfer. The DMA is initiated using a custom written
Linux kernel module, which makes received data accessible via a Linux
character device. The processor part of the Zynq SoC is running a
Petalinux Linux distribution \cite{petalinux} on both cores and hosts
a simple TCP server, that streams all data from the character device.

\subsection{Speed and data volume consideration}

The data volumes are shown at figure \ref{fig:Maximum-data-volume}.
The SPIROC produces \SIrange[ list-units = brackets , range-units = brackets]{0}{2.4}{\kilo\byte}
packets depending on how many of up to 16 events were stored in the
acquisition phase. The LED light calibration fills all memory cells
of all SPIROCs and uses the maximum of the data volume per readout
cycle and defines the minimum required bandwidth for the DAQ, which
is \SI{17}{\mega\byte} per LDA per readout cycle. This corresponds
to \SI{85}{\mega\byte\per\second} for the ILD geometry and accelerator
repetition frequency of \SI{5}{\hertz}. 

Data from all SPIROCs has to be readout within \SI{\sim198}{\milli\second}
before the next acquisition cycle starts. For 6 HBUs in a slab, which
corresponds to 24 SPIROCs in two shared readout chain buses, the minimum
transfer speed is \SI{1.2}{\mega\bit\per\second}. A \SI{1.67}{\mega\hertz}
readout speed is currently implemented. The SPIROC itself specifies
a \SI{5}{\mega\hertz} readout clock frequency limit. Similarly,
a DIF has to send up to \SI{180}{\kilo\byte} data chunks with a speed
at least \SI{7.2}{\mega\bit\per\second}.  The currently used serial
protocol uses a \SI{10}{\mega\hertz} clock, providing \SI{8.42}{\mega\bit\per\second}
bandwidth. 

The LDA needs to be able to supply continuously \SI{85}{\mega\byte\per\second}
over a GbE port. This requirement is close to the Ethernet bandwidth
limit during the LED calibration and might get replaced for the final
detector in order to provide bandwidth overhead. 

\begin{figure}
\begin{centering}
\includegraphics[height=6cm]{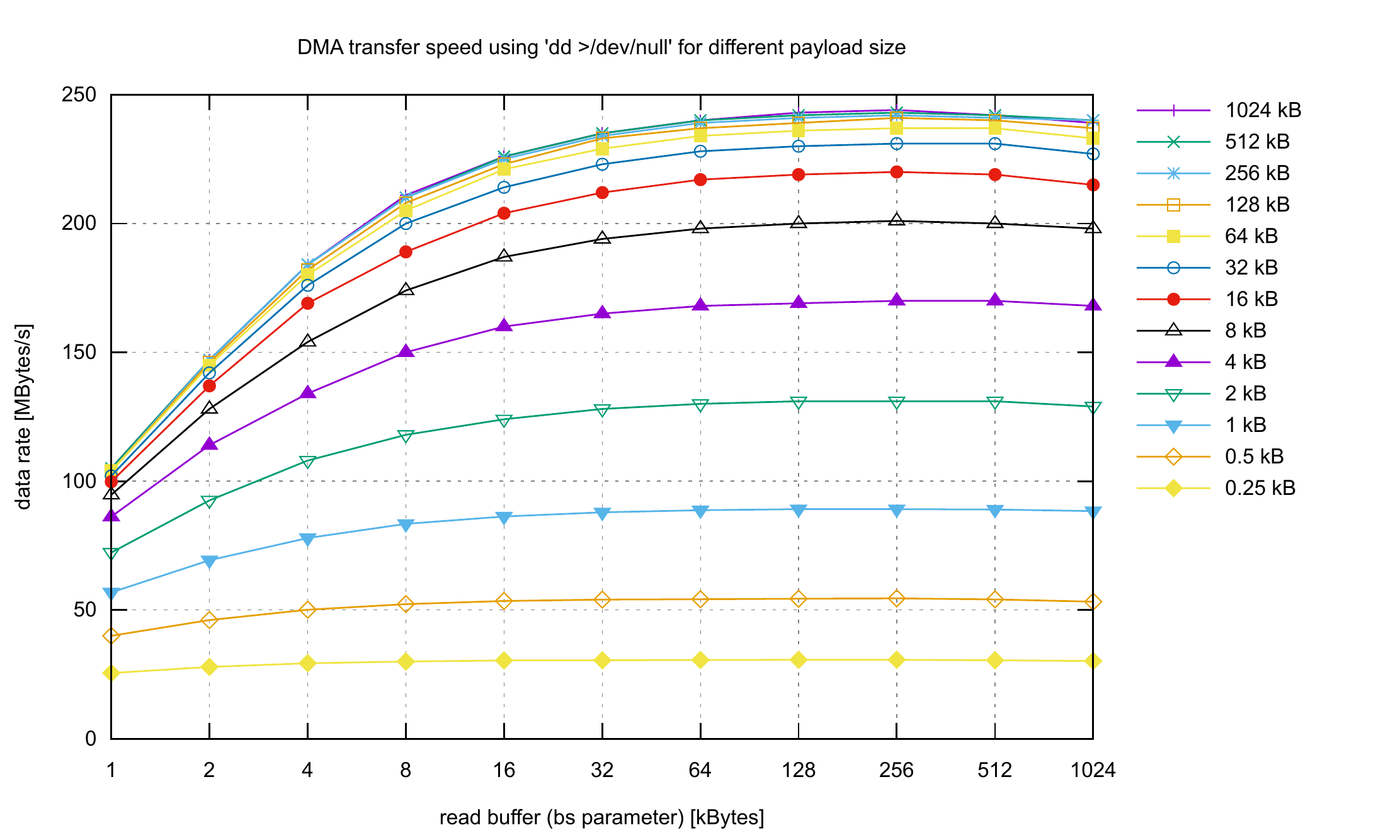}
\par\end{centering}
\caption{\label{fig:DMA-transfer-performance}DMA transfer data throughput}
\end{figure}

Since the LDA will be the device with the largest data traffic, it
has been subjected to bandwidth benchmarking. The FPGAs internally
process the data from the DIFs with \SI{320}{\mega\byte\per\second}.
The link between the FPGAs of a Wing-LDA is capable of transferring
nearly \SI{80}{\mega\byte\per\second} for each slave FPGA and the
further processing in the central FPGA has a \SI{320}{\mega\byte\per\second}
limit. The speed of the DMA transfer to the processor memory depends
on the size of the packets being moved by the DMA IP core and the
size of the buffer used for reading from the character device, as
shown in figure \ref{fig:DMA-transfer-performance}. The maximum transfer
speed of \SI{242}{\mega\byte\per\second} is achieved when a \SI{256}{\kilo\byte}
read buffer is used and small packets are merged into \SI{256}{\kilo\byte}
chunks in the FPGA. 

The software part in the SoC, which only sends the data from the memory
to the TCP socket, does not yet reach the requested bandwidth. The
maximum data transfer rate over the TCP was measured to be only \SI{6}{\mega\byte\per\second},
one order of magnitude below the requirements. 

It is not foreseen, that the current Linux TCP socket implementation
will be used in the final detector installation. It  will be replaced
for the final detector. The overall TCP performance is enough for
serving the beam tests with the currently available layers without
hitting the bandwidth limit, but will need to be optimized for larger
prototypes, especially when the spill rate will be higher than the
 \SI{5}{\hertz} expected at the ILC.

\section{\label{sec:Beam-test-operation}Beam test operation and  performance}

\begin{figure}
\begin{centering}
\includegraphics[width=0.6\textwidth]{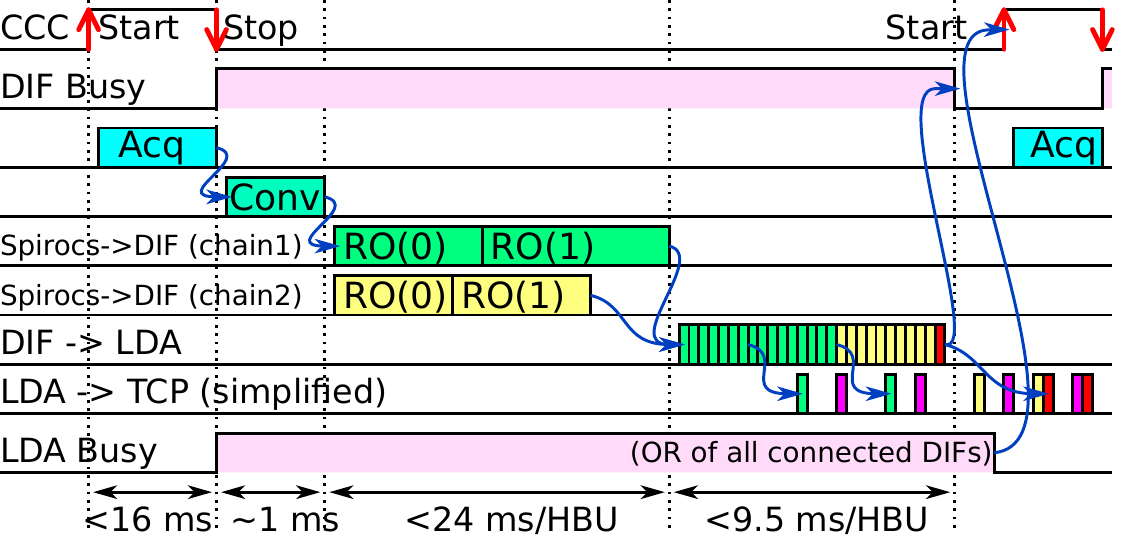}
\par\end{centering}
\caption{\label{fig:Beam-test-timing}Beam test timing showing the data transmission
sequence and busy signal operation. Acquisition (Acq) is followed
by the conversion (Conv), generating data stream, transferred over
2 readout channels per slab (only 4 asics in 2 readout chains are
shown). 100-bytes ASIC packet fragments are sent to LDA, followed
by the end-of-transfer packet (red). LDA sends packets from all DIFs
to PC (other DIF packets shown in violet).}
\end{figure}

Testbeam facilities have typically very different timing to the ILC
accelerator, which is expected to have \SI{1}{\milli\second} spills
with \SI{5}{\hertz} repetition rate, resulting in a \SI{0.5}{\percent}
duty cycle. However, testbeam facilities typically have a continuous
beam or rather long spills of continuous beam. Therefore, the following
measures are taken, leading to an operation shown in figure \ref{fig:Beam-test-timing}: 

\begin{figure}
\includegraphics[height=5cm]{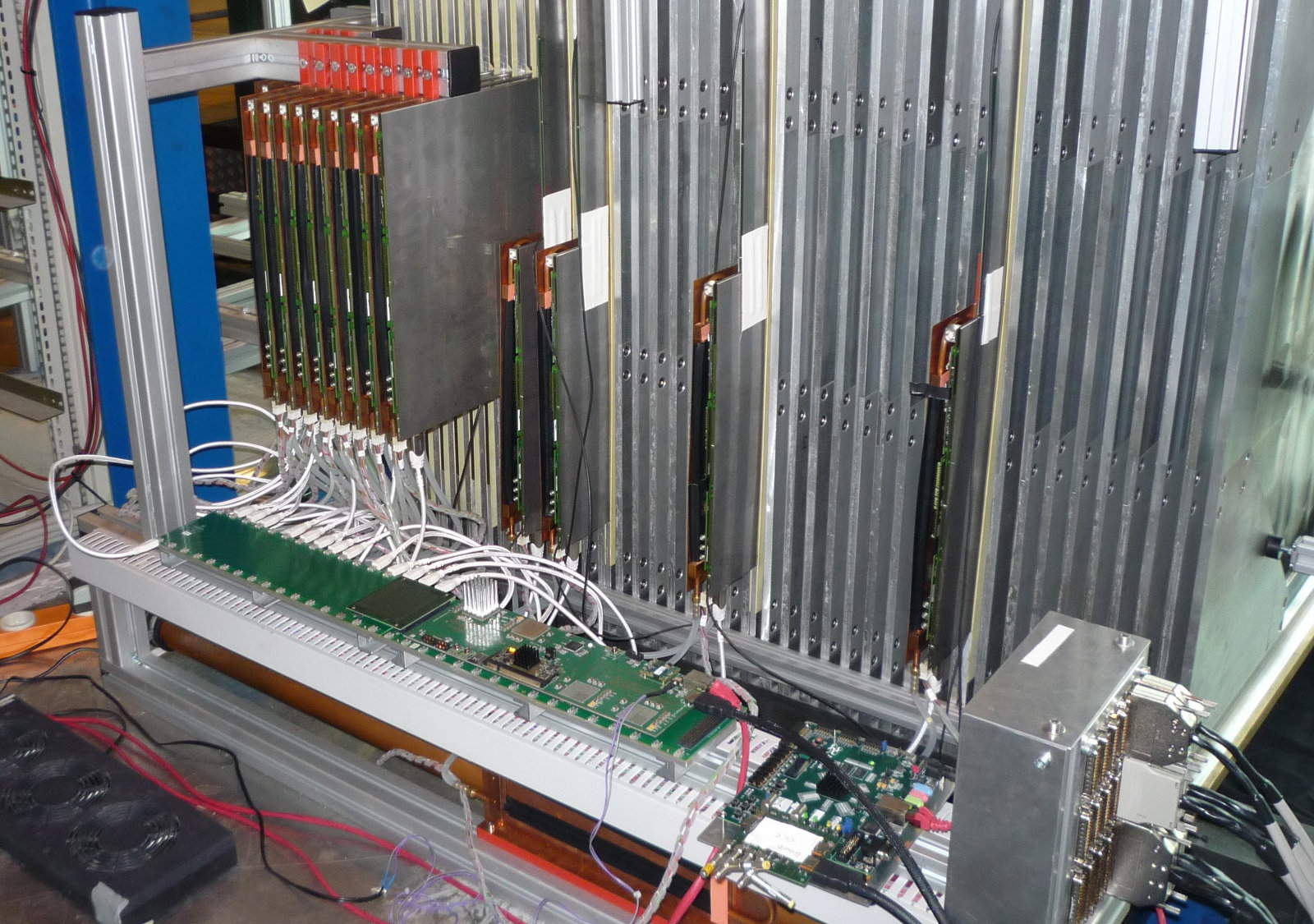}\hfill{}\includegraphics[height=5cm]{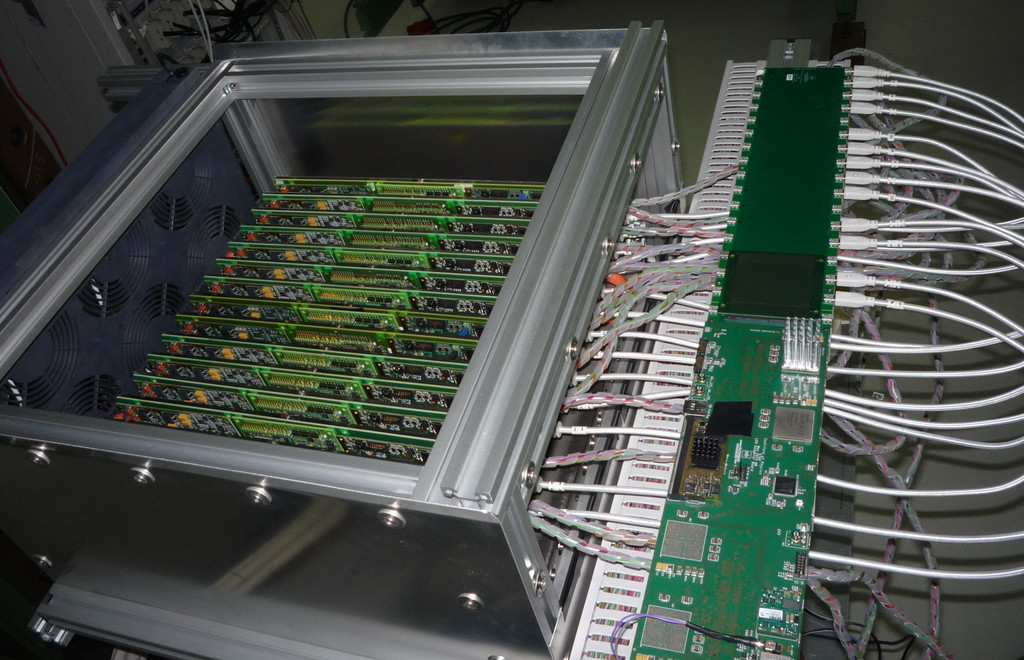}

\caption{\label{fig:testbeam-fotos}DAQ mechanical setup during beam tests
at CERN in 2015 with a tungsten stack (left), and at DESY in 2016
with a steel stack (right)}
\end{figure}

\begin{enumerate}
\item The bunch crossing period is extended from \SI{200}{\nano\second}
(close to \SI{\sim300}{\nano\second} of ILC) to \SI{4}{\micro\second},
prolonging the acquisition window up to \SI{16}{\milli\second}.
\item External trigger validation is introduced, otherwise SiPM noise would
become significant. Any auto-triggered event, which is not validated
within the \SI{4}{\micro\second} period is therefore not saved. This
validation scheme is not foreseen for ILC running.
\item A busy signal is raised when all 16 memory cells of a SPIROC in any
layer are filled, forcing all other layers to stop. The busy is propagated
through LDAs to CCC. The busy from DIFs is cleared when all data is
sent out.
\item Acquisition is restarted as soon as all data are sent to LDAs and
LDAs have enough free space in their buffers for the next acquisition
cycle.
\end{enumerate}

The DAQ system in the beam test configuration has been used since
the end of  2014 in many beam tests at CERN PS, CERN SPS and DESY
for different layer configurations of up to 15 layers. The Wing-LDA
connection to the detector layers is shown in figure \ref{fig:testbeam-fotos}.
The performance depends primarily on the number of HBUs in the readout
chain, since the transfer of data from SPIROC to DIF is the most limiting
factor.

The performance depends also on the beam condition and on SiPM noise,
which can be partially controlled by a trigger threshold and SiPM
bias voltage. The typical performances, that have been observed during
a recent beam test at DESY in 2016 are shown in table \ref{tab:performance}.
The acquisition duty cycle in testbeam mode is up to two orders of
magnitude higher than the designed value of \SI{0.5}{\percent}, providing
a possibility to efficiently record even a very low particle rate.
The system was able to record  up to 200~particles$\cdot\si{\per\second}$
 in the DESY test beam, which has a non-homogeneous particle burst
structure.

\begin{table}
\caption{\label{tab:performance}Performance during beam tests at DESY. ROC
= ReadOut Cycle.}

\centering{}%
\begin{tabular}{|c|c|c|c|}
\hline 
configuration & 2$\times$2 HBUs & 1$\times$1 HBU  & ILC (\SI{1}{\milli\second} spill)\tabularnewline
\hline 
layers & 4 & 15 & 15\tabularnewline
SPIROCs in a layer & 16 & 4 & 4\tabularnewline
ROC$\cdot$\si{\per\second} & 17 & $25\sim30$ & 99\tabularnewline
duty cycle & \SI{10}{\percent} & ($20\sim30$) \% & \SI{9.2}{\percent}\tabularnewline
particles/s & 135 & $150\sim200$ & 41\tabularnewline
\hline 
\end{tabular}
\end{table}

The ILC configuration was tested with a forced \SI{1}{\milli\second}
spill length maximum and \SI{200}{\nano\second} bunch crossing period.
The acquisition restarted as soon as possible, exceeding the expected
\SI{5}{\hertz} ILC spill rate by a factor of 20.

\section{Summary }

The AHCAL DAQ hardware was developed with the aim of serving beam
tests of the detector prototypes with the added feature of  being
scalable to a large ILC detector. The largest on-detector concentration
device, the Wing-LDA, has been stably serving this purpose since 2014
and is already scalable to the full size of AHCAL barrel, reading
out up to 8 million channels. 

The DAQ functionality was demonstrated in several beam tests with
up to 15 layers containing up to 108 ASIC and 3888 SiPM channels.
The performance showed 200 recorded particles per second during beam
tests and 450 calibrating light pulses per second during calibration.
The detector duty cycle can be increased by two orders of magnitudes
from the \SI{0.5}{\percent} ILD duty cycle due to DAQ testbeam mode
adaptations, which enable efficient data taking even in continuous,
low-intensity beams.

The current hardware is able to process internally up to \SI{320}{\mega\byte\per\second}.
The implementation of TCP server in the embedded Linux is sufficient
for beam test, but does not yet meet the bandwidth requirements for
the full-size detector and will have to be improved before testing
even a significantly larger ( $\sim3\times$) number of prototype
layers.

\appendix
\acknowledgments

The author gratefully thanks Mathew Wing, Katja Krüger, Mathias Reinecke,
Eldwan Brianne, Felix Sefkow and Jaroslav Cvach for their very useful
comments and feedback to this paper.

This project has received funding from the European Union\textquoteright s
Horizon 2020 Research and Innovation programme under Grant Agreement
no. 654168. 

This project has received funding from Ministry of Education, Youth
and Sports of the Czech Republic under the project LG14033.

\textcircled{c} for Figure \ref{fig:The-AHCAL} Mathias Reinecke
(DESY), used with permnission of author. 



\begin{thebibliography}{10}
\bibitem{calice}The CALICE collaboration. https://twiki.cern.ch/twiki/bin/view/CALICE/WebHome

\bibitem{pfa}Thomson, M. A., \emph{Particle flow calorimetry and
the PandoraPFA algorithm}. Nuclear Instruments and Methods in Physics
Research Section A: Accelerators, Spectrometers, Detectors and Associated
Equipment. Vol 611.1 (2009): 25-40

\bibitem{daq_concept}Wing, M. et al., \emph{A proposed DAQ system
for a calorimeter at the International Linear Collider}, LC note,
LC-DET-2006-008. 

\bibitem{DAQ_Goodrick}Goodrick, M.J. et al., \emph{Development of
a modular and scalable data acquisition system for calorimeters at
a linear collider}, JINST 6 (2011) P10011.

\bibitem{tdr_ild}Behnke, T. et al., \emph{The International Linear
Collider Technical Design Report-Volume 4: Detectors}. arXiv:1306.6329
(2013).

\bibitem{spiroc} Di Lorenzo, S. Conforti, et al., \emph{SPIROC: design
and performances of a dedicated very front-end electronics for an
ILC Analog Hadronic CALorimeter (AHCAL) prototype with SiPM readout}.
2013 JINST 8 C01027.

\bibitem{mark_engineering_prototype}Terwort, M., \emph{Concept and
status of the CALICE analog hadron calorimeter engineering prototype}.
Physics Procedia 37 (2012): 198-204.

\bibitem{xilinx_zynq}https://www.xilinx.com/products/silicon-devices/soc/zynq-7000.html

\bibitem{zedboard}http://zedboard.org/product/zedboard

\bibitem{mars-zx3}http://www.enclustra.com/en/products/system-on-chip-modules/mars-zx3/

\bibitem{axi-stream}\emph{AMBA AXI4-Stream Protocol Specification
v1.0}, www.arm.com

\bibitem{petalinux}http://www.xilinx.com/products/design-tools/embedded-software/petalinux-sdk.html
\end{thebibliography}
\end{document}